\documentclass{aa}  
\usepackage{graphicx}
\usepackage{txfonts}
\usepackage{lscape}
\usepackage{booktabs}
\usepackage{threeparttablex}
\usepackage{wasysym}
\usepackage{arydshln}
\usepackage{mathrsfs}
\usepackage{natbib}
\usepackage{amssymb}
\usepackage[bookmarks=false,colorlinks=true,linkcolor=black,citecolor=cyan,filecolor=black,urlcolor=cyan]{hyperref}

\begin{document} 

\title{Impact of the observation frequency coverage on the significance of a gravitational wave background detection in PTA data}

\author{
Irene Ferranti,$^{1,2}$\thanks{E-mail: irene.ferranti@unimib.it}
Mikel Falxa,$^{1}$\thanks{E-mail: mikel.falxa@unimib.it}
Alberto Sesana,$^{1,2,3}$\\
A. Chalumeau,$^{4}$
N. Porayko,$^{5}$
G. Shaifullah,$^{1,2,3}$
I. Cognard,$^{6, 7}$
L. Guillemot,$^{6, 7}$
M. Kramer,$^{5}$
K. Liu,$^{8}$
G. Theureau$^{6, 7, 9}$}

\date{Accepted XXX. Received YYY; in original form ZZZ}
\institute{
    Dipartimento di Fisica ``G. Occhialini", Universit{\'a} degli Studi di Milano-Bicocca, Piazza della Scienza 3, I-20126 Milano, Italy
    \and
    INFN, Sezione di Milano-Bicocca, Piazza della Scienza 3, I-20126 Milano, Italy
    \and
    INAF - Osservatorio Astronomico di Brera, via Brera 20, I-20121 Milano, Italy
    \and
    ASTRON, Netherlands Institute for Radio Astronomy, Oude Hoogeveensedijk 4, 7991 PD Dwingeloo, The Netherlands
    \and
    Max-Planck-Institut für Radioastronomie, Auf dem Hügel 69, 53121 Bonn, Germany
    \and
    LPC2E, OSUC, Univ Orleans, CNRS, CNES, Observatoire de Paris, F-45071 Orleans, France
    \and
    Observatoire Radioastronomique de Nan\c{c}ay, Observatoire de Paris, Universit\'e PSL, Université d'Orl\'eans, CNRS, 18330 Nan\c{c}ay, France
    \and
    LUTH, Observatoire de Paris, PSL Research University, CNRS, Universit\'e Paris Diderot, Sorbonne Paris Cit\'e, F-92195 Meudon, France
    }

\titlerunning{Impact of the observation frequency coverage on the significance of a GWB detection in PTA data}

\authorrunning{Ferranti, Falxa et al.}

\abstract{Pulsar Timing Array (PTA) collaborations gather high-precision timing measurements of pulsars with the aim of detecting gravitational wave (GW) signals. A major challenge lies in the identification and characterisation of the different sources of noise that may hamper their sensitivity to GWs. The presence of time-correlated noise that resembles the target signal might give rise to degeneracies that can directly impact the detection statistics. In this work, we focus on the covariance that exists between a "chromatic" dispersion measure (DM) noise and an "achromatic" stochastic gravitational wave background (GWB). "Chromatic" associated to the DM noise means that its amplitude depends on the frequency of the incoming pulsar photons measured by the radio-telescope. Several frequency channels are then required to accurately characterise its chromatic features and when the coverage of incoming frequency is poor, it becomes impossible to disentangle chromatic and achromatic noise contributions. In this paper we explore this situation by injecting realistic GWB into 100 realizations of two mock versions of the second data release (DR2) of the European PTA (EPTA), characterized by different frequency coverage. The first dataset is a faithful copy of DR2, in which the first half of the data is dominated by only one frequency channel of observation; the second one is identical except for a more homogeneous frequency coverage across the full dataset. We show that for 91$\%$ of the injections, a better frequency coverage leads to an improved statistical significance ($\approx$1.3dex higher log Bayes factor on average) of the GWB and a better characterisation of its properties. We propose a metric to quantify the degeneracy between DM and GWB parameters and show that it is correlated with a loss of significance for the recovered GWB and an increase in the GWB bias towards a higher and flatter spectral shape. In the second part of the paper, this correlation between the loss of GWB significance, the degeneracy between the DM and GWB parameters and the frequency coverage is further investigated using an analytical toy model.}
    

\maketitle


\section{Introduction}

Pulsar timing arrays \citep[PTAs,][]{1990ApJ...361..300F} exploit the exquisite rotational stability of millisecond pulsars to search for gravitational waves (GWs). Indeed, high precision timing measurement of millisecond pulsars provides us with datasets that are sensitive enough to measure perturbations of the curvature of space-time due to the passage of GWs at  nano-Hz frequencies. In this band, a stochastic GW background (GWB) is expected to be produced by a population of Supermassive black hole binaries \citep[SMBHBs,][]{1995ApJ...446..543R,Jaffe2003,2003ApJ...590..691W,2008MNRAS.390..192S}, or by a number of physical processes occurring in the Early Universe, from inflation to phase transitions, from scalar perturbations to domain walls and many more \citep[see][and references therein]{2018CQGra..35p3001C,2023ApJ...951L..11A, 2024A&A...685A..94E}. Last year, the European PTA (EPTA) collaboration  along with the Indian PTA (InPTA) collaboration  \citep[][]{wm3}, the North American Nanohertz Observatory for Gravitational Waves collaboration \citep[NANOGrav,][]{NANOGrav}, the Parkes PTA collaboration \citep[PPTA,][]{PPTA} and the Chinese PTA collaboration \citep[CPTA,][]{CPTA}, reported evidence for the presence of a GWB-like signal in their dataset at a 2-to-4$\sigma$ significance (depending on the dataset), effectively opening the nano-Hz GW sky.  

Pulsars are monitored in the radio-frequency band where a series of time of arrival (TOA) of pulses are extracted from the raw data, at given epochs. PTA datasets are made of timing residuals $\delta \vec{t}$ that are obtained by fitting a timing model (TM) to the observed TOAs of pulsars \citep[e.g.][]{tempo2, wm1}. The TM is able to predict the TOAs at the Solar system barycenter (SSB) by taking into account every physical process that occurs between emission and reception of the pulses and that can be deterministically modelled (i.e. pulsar spin-down, Einstein delay, Shapiro delay, ...). The timing residuals are the difference between the observed TOAs and the TM predicted TOAs. Any feature that is not modelled by the TM will still be present in $\vec{\delta t}$. These include stochastic processes that cannot be modelled by deterministic functions and, possibly, a GWB. Data analysis pipelines have thus been developed to simultaneously fit for putative GW signals and pulsar noises in the data  \citep{nanograv_noise_budget, wm2},

Extracting a GW signal from PTA data and properly assessing its significance is no easy task, as pulsar observations suffer from a myriad of subtle stochastic noise sources \citep[e.g.][]{2013CQGra..30v4002C,2016MNRAS.455.4339T}. Since a GWB produces a stochastic time correlated signal (i.e. a stochastic red signal), any noise with the same statistical properties can blend with it, affecting its recovery. The two main components of time-correlated noise are (i) the red noise (RN): due to stochastic variations of the pulsar spin rate \citep[e.g.][]{2010ApJ...725.1607S} (ii) the dispersion measure (DM) noise: due to stochastic variations of the electron density in the interstellar medium, interacting with the pulses along the line of sight \citep[e.g.][]{2016ApJ...817...16C}. Each of these physical phenomena can be recognized in PTA data by its peculiar properties. The GWB is common to all pulsars, is achromatic (i.e. it does not depend on the observing radio frequency), and feature a quadrupolar spatial correlation pattern depending on the pulsars' angular separation \citep{hd1983}. RN is also achromatic, but its spectrum can vary from pulsar to pulsar and no inter-pulsars spatial correlation is expected. DM noise is also expected to be specific to each pulsar with no spatial correlations, but it is chromatic, with an amplitude that is square inversely  proportional to the frequency $\nu$ of the observed photon \citep[e.g.][]{2007MNRAS.378..493Y}. Therefore, the three contribution can be in principle easily separated by a PTA with a large number of high quality pulsars, uniformly distributed in the sky and monitored in a wide radio-frequency band. Unfortunately, none of these conditions is fully met by current PTAs: current GWB evidence is dominated by just a handful of pulsars, and observation frequency coverage is scant, especially for old data.

DM noise can be particularly problematic since its variation can be orders of magnitude higher than that of the GWB and it typically affects all the pulsars. Failure in modelling it properly can therefore cause significant leakage into RN and GWB, eventually compromising the GWB recovery.
In order to characterise the DM noise, multi-frequency observations are required. This is the main reason why PTAs have been striving to expand the frequency coverage of their receivers and to devise new techinques to produce TOAs from wide band observations  \citep{2014MNRAS.443.3752L,2014ApJ...790...93P}.  
Since the observations are performed using different radio-frequency channels, the timing residuals can be divided into several observation frequencies. In that sense, a PTA dataset is not only a one-dimensional data stream, but rather a two-dimensional surface of timing residuals on the frequency $\nu$ and TOA $\Vec{t}$ plane. However, as mentioned above, this surface is often not uniformly covered, as it was the case for the second EPTA data release \citep[EPTA DR2,][]{wm1}, in which the first half of the data is dominated by essentially only one observation frequency. This can introduce degeneracies between different noise components, corrupting the recovery of the GWB. Indeed, this might be one of the reasons why the GWB evidence in EPTA DR2 drops when also the first half of the data is included in the analysis \citep{wm3}, despite the fact that we expect the GWB evidence to increase with increasing observation time.

It is therefore of primary interest to study the relation between observation frequency coverage, DM mis-modeling and GWB detection and parameter estimation. To this end, in this paper we construct two sets of synthetic PTA datasets mimicking the 24.8 years of observations collected in the EPTA DR2. One set is a faithful copy of DR2, featuring a single frequency channel in the first half of the data; the second one is identical except for the fact that observations are evenly distributed across the different frequency channels. We inject a realistic mock GWB from a cosmic population of SMBHBs \citep[e.g.][]{Rosado_2015} and make use of standard Bayesian analysis techniques to extract the signal from the data. We demonstrate that the absence of adequate frequency coverage is detrimental to GWB recovery and parameter estimation, which has important implications for interpreting the EPTA DR2 results \citep{wm3}, for forecasting PTA sensitivities \citep{Rosado_2015,2016ApJ...819L...6T,2023MNRAS.518.1802S}, and for planning future PTA observations \citep[e.g.][]{2012MNRAS.423.2642L,2018ApJ...868...33L}. 

The paper is organized as follows. In Section \ref{sec:sims}, we present the EPTA data \citep{wm1, wm2}, we explain how realistic simulations of it featuring different levels of frequency coverage are generated, and we introduce the models and statistical tools used in the subsequent analysis. In Section \ref{sec:res}, using these realistic simulations, we show the existence of a significant correlation between the DM-GWB and RN-GWB degeneracy and the GWB evidence and parameter estimation, and we present an analytical toy model that qualitatively explains the observed correlations. Finally, we summarize our main findings and draw our conclusions in Section \ref{sec:con}.

\section{Simulations and analysis tools}
\label{sec:sims}

\begin{figure*}
   \centering
   \includegraphics[width=1\textwidth]{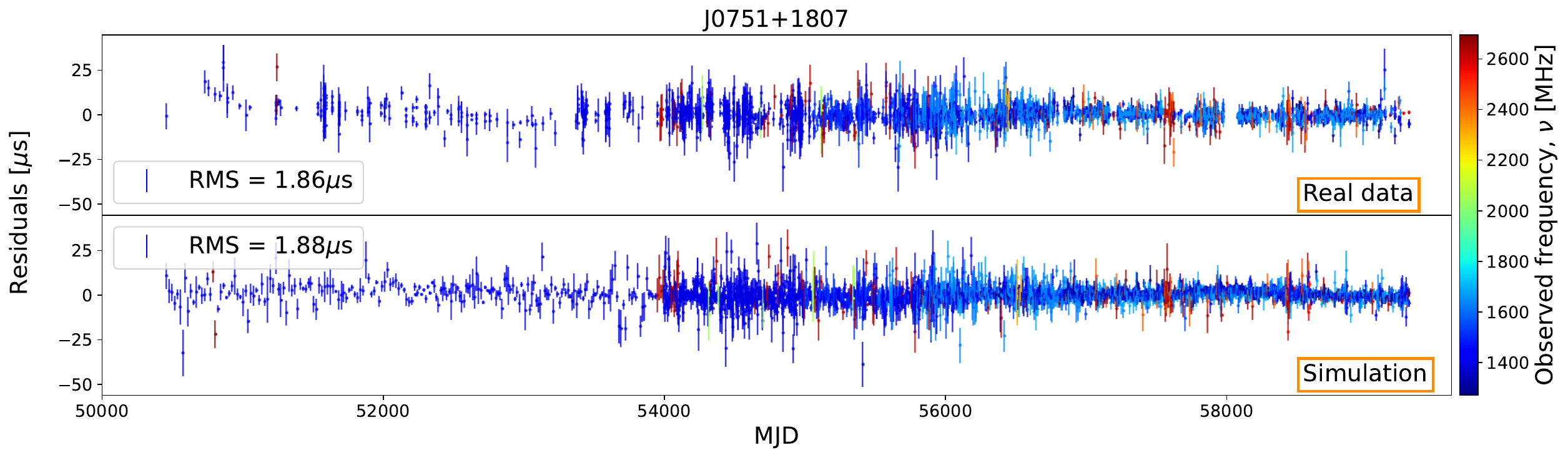}
   \caption{Comparison between real data from EPTA-DR2full (top panel) and simulated data (bottom panel) for pulsar J0751+1807.}
   \label{fig:residuals}%
\end{figure*}

\begin{figure*}
   \centering
   \includegraphics[width=1\textwidth]{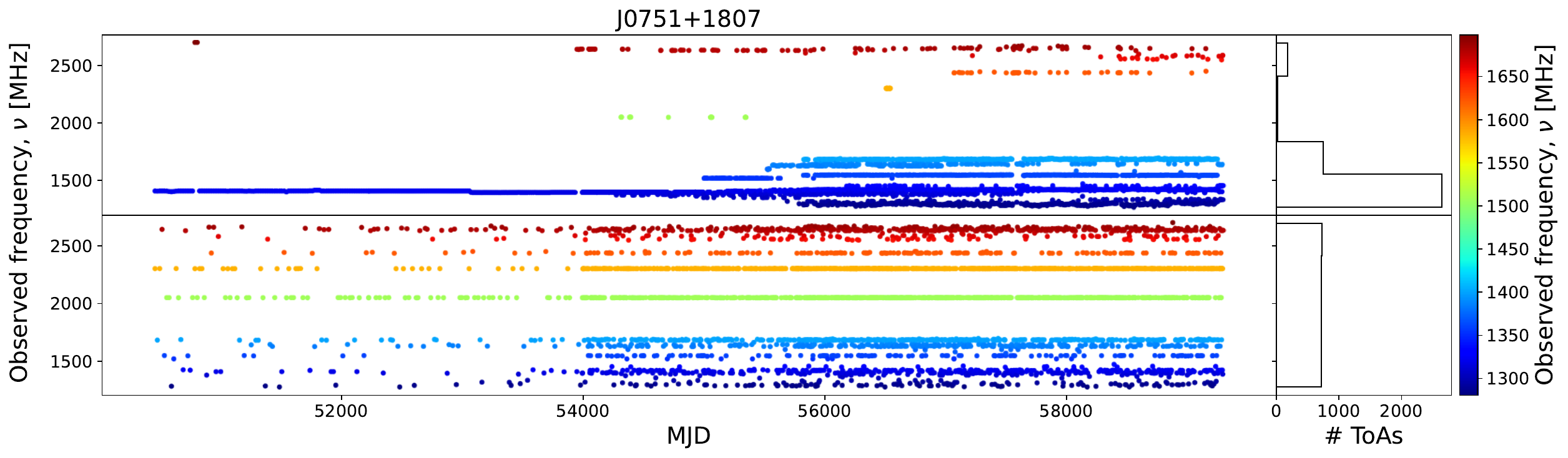}
   \caption{Frequency distribution of the times of arrival as a function of the observation epochs. The top panel shows the simulated DR2full-like dataset, the bottom panel shows the DR2FC-like (same epochs as DR2full, but with homogeneous distribution of observation frequencies among epochs) dataset. The histograms on the right show the distribution of the number of observations across frequencies.}
   \label{fig:FrequencyCoverage}%
\end{figure*}

We investigate the impact of the observation frequency coverage on the detection significance and parameter estimation of the observed GWB by means of realistic simulations of PTA datasets. The simulations intend to copy the real EPTA DR2 dataset by using the same epochs TOAs, observation frequencies and noise properties as described in \cite{wm1, wm2}. We generate empty pulsars with only white noise and inject stationary RN and DM noise. Then, a realistic GW background is added by injecting a large number ($\sim$120k) of individual GW sources taken from SMBHB population simulations.

\subsection{Reference real dataset}

The dataset we took as a reference is the EPTA \texttt{DR2full}, which consists of 24.8 years of observation \citep{wm1}, 25 pulsars, stationary noise, RN and DM. EPTA data are taken from the major European radio telescopes: the Effelsberg 100-m radio telescope (EFF) in Germany, the 76-m Lovell Telescope and the Mark II Telescope at Jodrell Bank Observatory (JBO) in the United Kingdom, the large radio telescope operated by the Nançay Radio Observatory (NRT) in France, the 64-m Sardinia Radio Telescope (SRT) operated by the Italian National Institute for Astrophysics (INAF) through the Astronomical Observatory of Cagliari (OAC), and the Westerbork Synthesis Radio Telescope (WSRT) operated by ASTRON, the Netherlands Institute for Radio Astronomy. These telescopes also operate together as the Large European Array for Pulsars (LEAP), which gives an equivalent diameter of up to 194m \citep[]{2016MNRAS.460.2207B}. Contemporary measure from multiple telescopes allows us to have multi-frequency observations in a large frequency band, covering the range 323MHz-4857MHz. However, this large frequency band is mainly present in the second half of the data, while the first 10-15 years of observations are strongly dominated by measurements in only one narrowband, around 1400MHz. Therefore, the frequency coverage is very inhomogeneous, specially on the pulsars with the longest observation time spans.

\subsection{Simulation of a realistic PTA dataset}\label{sec:dr2_sim}

Here we give a brief description of how the simulated datasets are produced. We simulated two copies of the EPTA \texttt{DR2full} dataset: the first reproduces the same frequency channel distribution as the real dataset, the second has an even distribution of the radio frequency channels across the entire time span of observation. The simulation pipeline starts by reproducing the main properties of the observation epochs (the observation time, the number of observations and the cadence) of each pulsar, then it assigns an observation frequency to each of them. We generate:

\begin{itemize}
    \item \texttt{DR2full}, where we assign to each simulated epoch the same radio frequency that appears in the real dataset;
    \item \texttt{DR2FC}, where we redistribute the observation frequencies at each epoch. The new radio frequencies are uniformly chosen among the actually observed frequencies for that pulsar.
\end{itemize}

Together with the observation frequency, also the corresponding observatory, backend and timing error are associated to the epoch\footnote{Since every backend has its characteristic level of white noise, changing the number of the epochs per backends changes the total RMS of the pulsars' residuals. However, on average, the RMS stays within 1.2 times the value of \texttt{DR2full} RMS.}. Data are then generated using the \texttt{libstempo} package \citep{2020ascl.soft02017V}, assuming the timing model in \cite{wm1}. The same package is used to inject the three components of the noise budget: white noise, red noise and DM variations, as we describe next.

\subsection{Simulating stationary Gaussian noise}
\label{subsec:stat_gp_noise}
The three noise components used in this paper are White Noise (WN), Red Noise and Dispersion measures. They are all stationary Gaussian processes, with the difference that WN is not correlated in time, while RN and DM are time-correlated signals. All noise processes $n(t)$ are described in the time domain by their covariance matrix $\langle n(t_i) n(t_j) \rangle$. Since white noise is uncorrelated in time, its covariance matrix is diagonal, with the elements given by
\begin{equation}\label{wn}
    \sigma_i = \sqrt{EFAC_i^2\sigma_{\rm ToA}^2 + EQUAD_i^2},
\end{equation}
where $\sigma_{\rm ToA}$ is the errorbar associated with each ToA according to the template fitting errors, EFAC accounts for the ToA measurement errors, and EQUAD accounts for any putative additional source of white noise. EFAC and EQUAD are specific to each observing backend. This model simply means that ToAs affected by the WN process are Gaussian distributed around the value predicted by the TM, where $\sigma_i$ is the width of the Gaussian.\\
Time-correlated signals are modelled as a combination of sine and cosine basis terms weighted by normally distributed coefficients \citep{vv2014}: for a noise $n(t)$ evaluated at time $t_i$ with power spectral density (PSD) $S(f)$ evaluated at frequencies $f_k$, we have
\begin{equation}
    n(t_i) = \sum_k \sqrt{S(f_k) \Delta f_k} \bigg (\frac{\nu_{\rm ref}}{\nu_i} \bigg )^{id} \bigg(X_k \sin(2\pi f_k t_i) + Y_k \cos(2\pi f_k t_i) \bigg)
\end{equation}
with $X_k, Y_k \sim \mathcal{N}(0, 1)$, $\Delta f_k = f_{k+1} - f_k$, $\nu_{\rm ref}$ a reference frequency (here set to 1.4 GHz), $\nu_i$ the observation frequency corresponding to $t_i$ and $id$ the chromatic index. To generate one realisation of the noise $n$, we randomly draw coefficients $X_k, Y_k$ from their normal distribution in order to obtain the Fourier coefficients associated with frequency $f_k$. The covariance matrix corresponding to this kind of noise process is
\begin{equation}
\begin{aligned}
    \langle n(t_i) n(t_j) \rangle = \sum_k & S(f_k) \Delta f \bigg (\frac{\nu_{\rm ref}}{\nu_i} \bigg )^{id} \bigg(\frac{\nu_{\rm ref}}{\nu_j}  \bigg )^{id} \\
    & \times \cos(2\pi f_k (t_i - t_j))
\end{aligned}
\label{eq:cov}
\end{equation}

Where the chromatic index $id$ determines the frequency dependence of the noise. For achromatic noise $id=0$, whereas for DM noise $id=2$.\\




In Fig. \ref{fig:residuals}, a comparison between the real and the simulated data is shown for pulsar PSR J0751+1807. In Fig. \ref{fig:FrequencyCoverage}, a comparison between the frequency coverage of \texttt{DR2full} and \texttt{DR2FC} is shown for pulsar PSR J0751+1807.

\subsection{Simulation of a realistic GWB}

To simulate an astrophysical GWB produced by SMBHBs, the injection pipeline starts from a population of circular SMBHBs, computes the variations induced by each binary in each pulsar ToAs and then sums these delays in the time domain. A SMBHBs population is characterised by the number of emitting sources per unit redshift, mass and frequency ${\rm d}^3N/({\rm d}z{\rm d}\mathcal{M}{\rm d}f_r)$, therefore a list of binaries is constructed as a random sample from the numerical distribution ${\rm d}^3N/({\rm d}z{\rm d}\mathcal{M}{\rm d}f_r)$, which is obtained from the observation-based models described in \cite{Sesana_2013} and \cite{Rosado_2015}. Through this method, hundreds of thousands of SMBHBs populations can be obtained, that are in agreement with current constraints on the galaxy merger rate, the relation between SMBHs and their hosts, the efficiency of SMBH coalescence and the accretion processes following galaxy mergers. The power spectrum produced by a population of SMBHBs can be computed as \citep{Rosado_2015}
\begin{equation}\label{eq:spectrum}
    h_{c}^{2}(f_i) = \sum_{j\in \Delta f_i}\frac{h_j^2(f_r)f_r}{\Delta f_i}
\end{equation}
where $f_i$ is the central frequency of the bin $\Delta f_i$, and the sum runs over all the binaries for which the observed frequency $f_r/(1+z)\in \Delta f_i$, where $f_r$ is the GW frequency in the binary rest frame and therefore $f_r/(1+z)$ is the observed frequency. $h_j(f_r)$ is the strain of each GW signal, given by
\begin{equation}
    h(f_r) = 2\sqrt{\frac{1}{2}\bigl(a(\imath)^2 + b(\imath)^2\bigr)}\frac{(G\mathcal{M})^{5/3}(\pi f_r)^{2/3}}{c^4d(z)},
\end{equation} 
where $\mathcal{M}$ is the binary chirp mass, $\imath$ the orbit inclination angle, $a(\imath) = 1+\text{cos}^2\imath$, $b(\imath) = -2\text{cos}\imath$ and $d(z)$ is the comoving distance of the source, which is computed as 
\begin{equation}
    d(z) = \frac{c}{H_0}\int_{0}^{z}\frac{1}{\sqrt{\Omega_m (1+z)^3 + \Omega_{\Lambda}}}{\rm d}z,
\end{equation}
where the cosmology assumed is $\Lambda$ cold dark matter and the set of cosmological parameters chosen is $(H_0, \Omega_m, \Omega_{\Lambda})=(70{\rm km s^{-1}Mpc^{-1}}, 0.3, 0.7)$.\\
Since the binaries considered here are circular, their GW emission is fully described by 8 parameters. $\mathcal{M}$, $f_r$ and $z$ are generated according to the observationally-based model described above, while inclination angle $\imath$, polarization angle $\psi$, initial phase $\Phi_0$ and sky location $(\theta, \phi)$ are randomly generated from uninformative distributions: $\text{cos}\imath \in \text{Uniform}(-1, 1)$, $\psi \in \text{Uniform}(0, \pi)$, $\phi_0 \in \text{Uniform}(0, 2\pi)$, $\text{cos}\theta \in \text{Uniform}(-1, 1)$ and $\phi \in \text{Uniform}(0, 2\pi)$.\\
Once these 8 parameters have been generated for each binary in the population, the variations induced in each pulsar are evaluated according to \cite{Ellis_2013}, taking into account both the Earth and pulsar terms. They are then all summed to give the total delay induced in each pulsar by the GWB.\\

\begin{figure}
   \centering
   \includegraphics[width=0.5\textwidth]{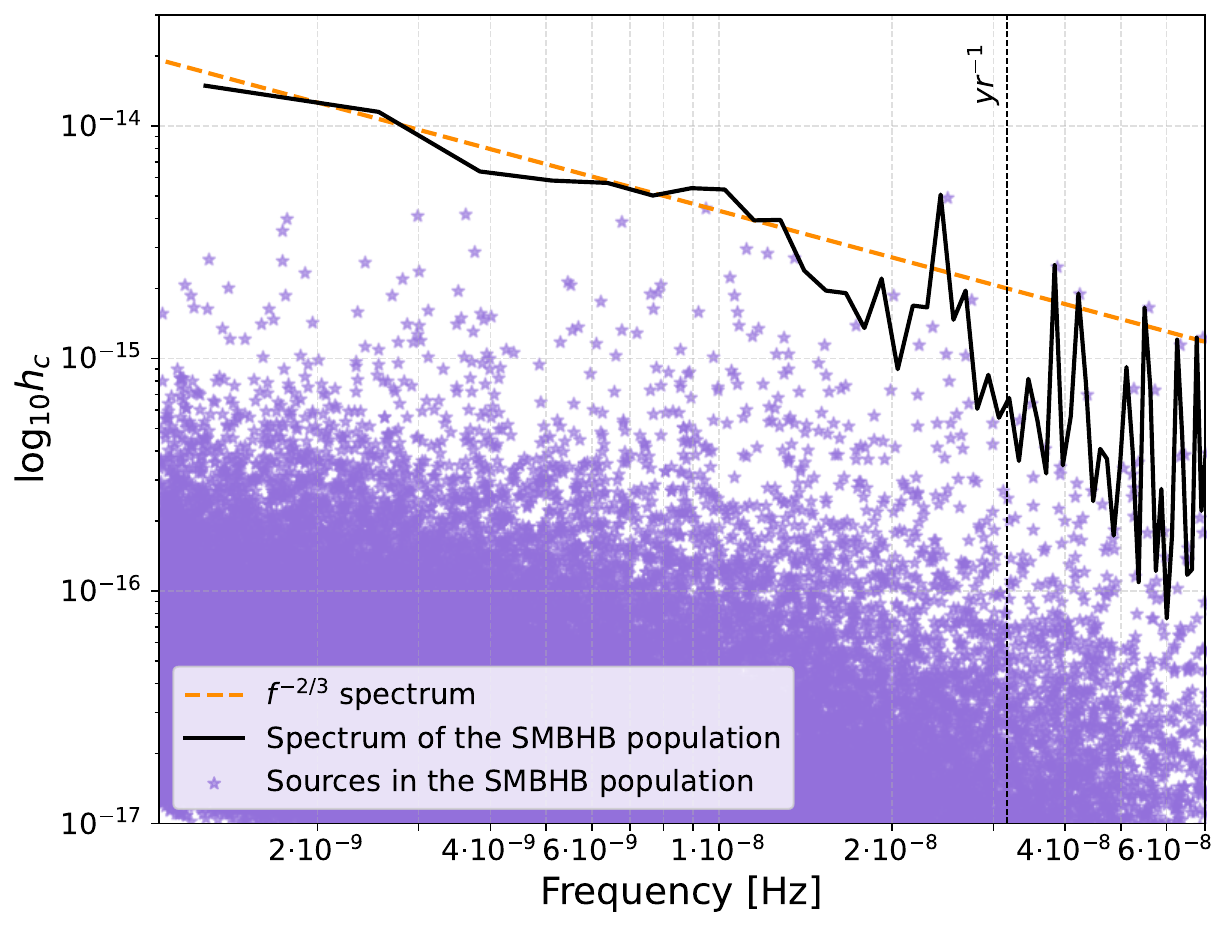}
   \caption{Characteristic strain spectrum of the injected GWB. Purple stars are the contributions to the spectrum of all the SMBHBs in the populations. The black solid line is the total power spectrum in each frequency bin, where the frequency bins start at $f=1/T_{\rm obs}$ and have width $\Delta f=1/T_{\rm obs}$ with $T_{\rm obs}=24.8$yr. As reference, the ideal spectrum with slope $-2/3$ and amplitude at the reference frequency of $1$yr$^{-1}$ $A_{\rm GW}=2.4\times 10^{-15}$ is shown by the dashed orange line.}
   \label{fig:GWB_spectrum}%
\end{figure}

For the purpose of this paper, only one GWB realization is used, which corresponds to the spectrum shown in \autoref{fig:GWB_spectrum}. The GWB realization is chosen such that the spectrum, if expressed in the power-law form
\begin{equation}\label{eq:hc_phinney}
    h_{c}(f) = A_{\rm GW}\left(\frac{f}{{\rm yr}^{-1}} \right)^{\alpha},
\end{equation}
has a slope $\alpha \sim -2/3$ at the lowest frequencies, which is the value expected for an ideal GW-driven population of circular binaries, and a nominal amplitude of $\sim 2.4\times10^{-15}$, which is consistent with the value inferred from EPTA DR2new \citep{wm3}.

\subsection{Data analysis}
\begin{figure*}
   \centering
   \includegraphics[width=1\textwidth]{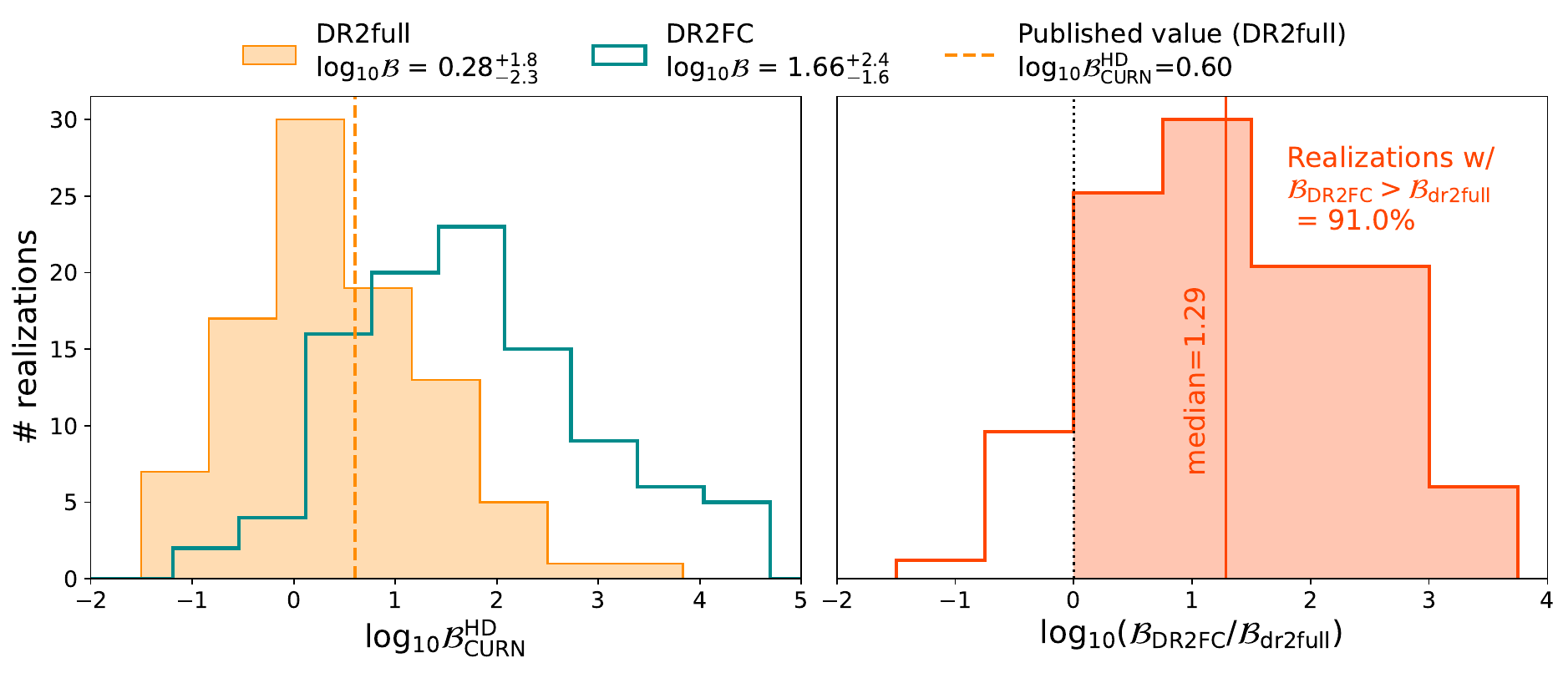}
   \caption{Left panel: distribution of log$_{10}\mathcal{B}^{\rm HD}_{\rm CURN}$ obtained with the 100 realizations of \texttt{DR2full}-like datasets (orange) and \texttt{DR2FC}-like datasets (green). The dashed orange line is the value of log$_{10}\mathcal{B}^{\rm HD}_{\rm CURN}$ obtained by the EPTA analysis of the real \texttt{DR2full} \citep{wm3}. Right panel: distribution of the ratio between $\mathcal{B}^{\rm HD}_{\rm CURN}$ obtained with \texttt{DR2full} and with \texttt{DR2FC}.}
   \label{fig:BF_fc}%
\end{figure*}
When analizing the data, we assume Gaussian and stationary noise and use a Gaussian likelihood. In PTA data analysis, it is custom to marginalise over first order errors of TM parameters \citep{vv2014}. The marginalisation is performed analytically and gives a TM marginalised expression of the likelihood of the form
\begin{equation}
    \ln \mathcal{L}(\delta t|\Vec{\theta}) \propto -\frac{1}{2} (\delta t - s)^\top C^{-1} (\delta t - s) - \frac{1}{2} \det {C},
\label{eq:likelihood}
\end{equation}
where $\delta t = \bigcup _{a=1} ^{N_{\rm PSR}} \delta \Vec{t}_a$, and $C = C_a \delta_{ab} + C_h \Gamma_{ab}$ is the covariance matrix. In the latter, $C_h$ characterises the achromatic common RN, spatially correlated between pulsars following the Hellings-Downs correlation pattern \citep{hd1983}

\begin{equation}
\Gamma_{ab}=\frac{1}{3}-\frac{1}{6}\frac{1-\text{cos}\gamma_{ab}}{2}+\frac{1-\text{cos}\gamma_{ab}}{2}\text{ln}\biggl(\frac{1-\text{cos}\gamma_{ab}}{2}\biggr),
\label{eq:hd}
\end{equation}
with $\gamma_{ab}$ the angular separation between the pulsars in the pair $ab$. $C_a$ represents the noise present in pulsar $a$ and can be written as

\begin{equation}
    C_a (t_i, t_j) = \sigma^2_i \delta _{ij} + C_{\rm RN} (t_i - t_j) + C_{\rm DM} (t_i - t_j),
\end{equation}
where $\sigma_i$ is the level of white noise at $t_i$, $C_{\rm RN}$ is the achromatic RN and $C_{\rm DM}$ the DM noise. $C_{\rm RN}$ and $C_{\rm DM}$ are given by \autoref{eq:cov}, respectively with $id=0$ and $id=2$. It is usually assumed that these two noise components have powerlaw spectra \citep{wm2} with amplitude and spectral index that have to be characterised for each pulsar. This complicates the task of fitting for a GWB since we also expect the latter to follow a powerlaw (see \autoref{eq:hc_phinney}). Thus, the covariance between noise and signal can be significant.

When performing Bayesian analysis, we update our prior knowledge on parameters by multiplying the likelihood with the prior probability distribution $\pi(\vec{\theta})$ for our parameters $\vec{\theta}$. The resulting probability is the posterior probability $p(\vec{\theta}|\delta t)$

\begin{equation}
    p(\vec{\theta}|\delta t) = \frac{\mathcal{L}(\delta t|\Vec{\theta})\pi(\vec{\theta})}{\int \mathcal{L}(\delta t|\Vec{\theta})\pi(\vec{\theta})d\theta}
\end{equation}
expressing the probability of measuring $\vec{\theta}$ given the timing data $\delta t$. The normalization factor $\int\mathcal{L}(\delta t|\Vec{\theta})\pi(\vec{\theta})d\theta$ is the Bayesian evidence, which is used to perform model comparison: the Bayes factor ($\mathcal{B}$) between two models 1 and 2 is given by the evidence of model 1 divided by the evidence of model 2:
\begin{equation}
    \mathcal{B}^1_2 = \frac{\mathcal{Z}_1}{\mathcal{Z}_2} = \frac{P(D|M_1)}{P(D|M_2)} ) = \frac{\int P(D|\theta_1, M_1)P(\theta_1, M_1)d\theta_1}{\int P(D|\theta_2, M_2)P(\theta_2, M_2)d\theta_2}.
\end{equation}
For the purpose of the analysis presented here, model 1 corresponds to an HD-correlated common red noise (which is the signature of a background of gravitational origin) and model 2 to a common, but uncorrelated, red noise.

\section{Results}
\label{sec:res}

\subsection{Comparing frequency coverages}

In this section, we show the results of the analyses performed on the two realistic datasets described in Section \ref{sec:dr2_sim}: (i) \texttt{DR2full}, which is a copy of the real EPTA dataset and (ii) \texttt{DR2FC}, which has the same observation epochs as \texttt{DR2full} but with a homogeneous coverage of the observation frequencies across the whole time of observation $T_{\rm obs}$. A comparison between simulated \texttt{DR2full} and \texttt{DR2FC} frequency coverage is shown in \autoref{fig:FrequencyCoverage} for pulsar J0751+1807. For both cases, we generated 100 datasets, each containing the same GWB signal (whose spectrum is shown in \autoref{fig:GWB_spectrum}), but featuring 100 different noise realizations.

\subsubsection{Comparison of Bayes Factor}

For the two types of datasets, we compute the Bayes Factor that compares a model with a common uncorrelated red noise (CURN) and an HD-correlated common red noise (HD) through the product space method implemented in \texttt{ENTERPRISE\_EXTENSIONS}. Together with the posterior distributions of the common red noise parameters (which are the slope $\gamma$ and the amplitude $A$ of the power spectrum), the parameters of individual pulsars red noise and DM noise are sampled. White noise parameters are fixed to the injected values. Therefore, the dimensionality of the posterior distribution is 64, including 62 noise parameters and 2 GWB signal model parameters.

In the left panel of \autoref{fig:BF_fc}, the distribution of $\mathcal{B}^{\rm HD}_{\rm CURN}$ obtained from the 100 realizations of noise is shown for \texttt{DR2full} and \texttt{DR2FC}.
Let us highlight the fact that the value of $\mathcal{B}^{\rm HD}_{\rm CURN}$ obtained with the real \texttt{DR2full} is 4 \citep{wm3}, which is well within the 68$\%$ C.I. of the distribution obtained with the realistic \texttt{DR2full} simulations, $\mathcal{B}^{\rm HD}_{\rm CURN} = 1.9^{+19.1}_{-1.7}$. This is an indication that the simulations are capturing the main features of the real data.

Focusing now on the comparison between the two datasets, we can conclude that the case with the better frequency coverage -- \texttt{DR2FC} -- gives significantly higher Bayes factors with respect to \texttt{DR2full} datasets. Indeed, in 91$\%$ of the cases \texttt{DR2FC} gives better results than \texttt{DR2full}, with the same noise realization. The average ratio between the \texttt{DR2FC} and \texttt{DR2full} Bayes factor is $\sim 20$, thus the average enhancement of the significance of signal recovery is more than one order of magnitude (see left panel of \autoref{fig:BF_fc}).\\


\subsubsection{Correlation between BF and constraints on the noise parameters}\label{BF_JSD}

\begin{figure}
   \centering
   \includegraphics[width=0.5\textwidth]{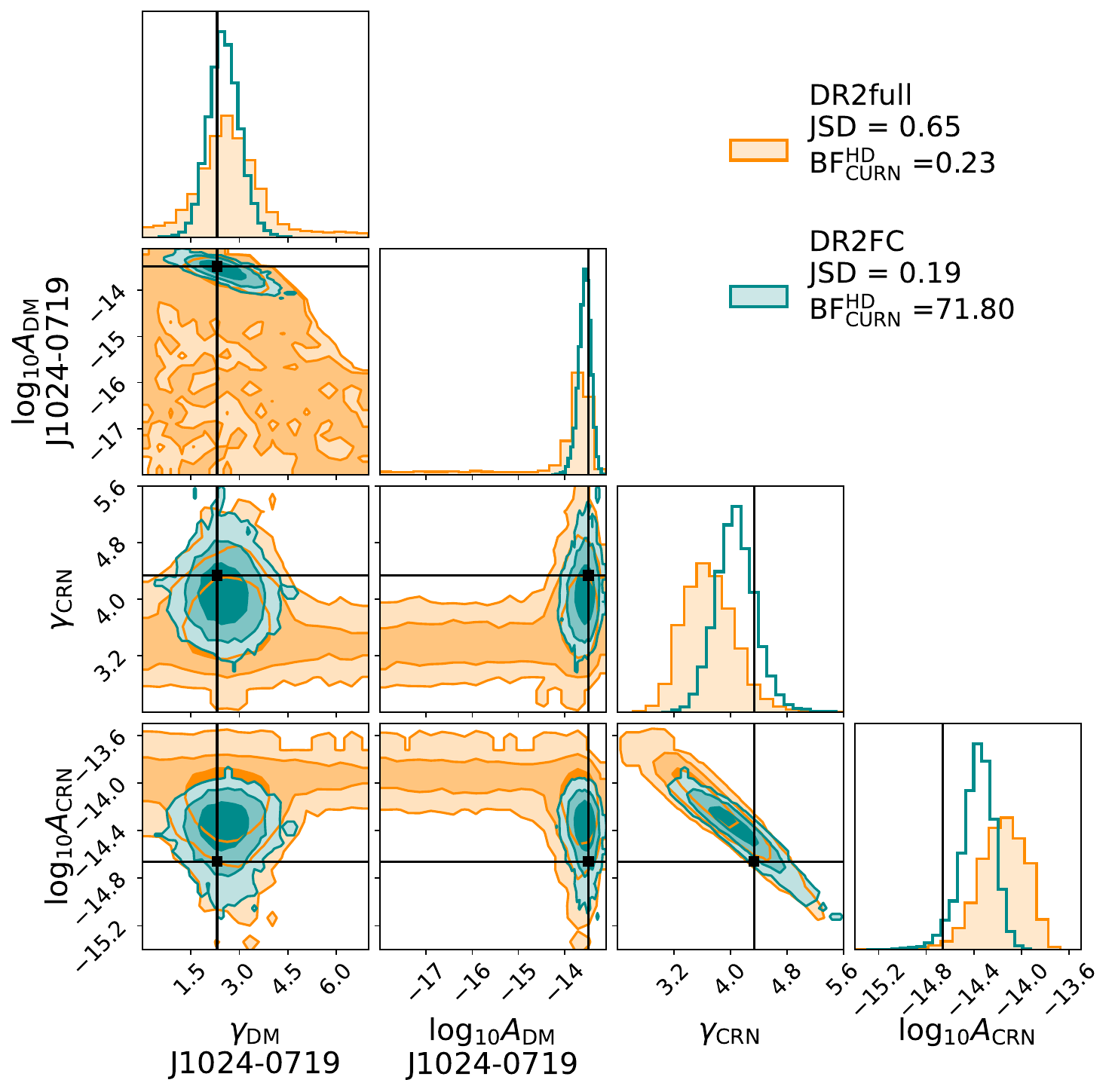}
   \caption{Posterior distributions of pulsar $J1024-0719$ individual noise and common red noise parameters. Distributions obtained with one realization of the \texttt{DR2full}-like dataset are shown in orange, while distributions obtained with the same realization of the \texttt{DR2FC}-like dataset are shown in green. Black squares are the injected values. JSD values and Bayes factors for both datasets are reported in the top right (see text for full description).}
   \label{fig:Lshape}%
\end{figure}

In this section, we investigate a possible reason for the increase in the Bayes factor due to the better frequency coverage. It is expected that a better frequency coverage improves the recovery of the dispersion measure parameters and helps to more easily disentangle this process from the achromatic red noise. Indeed, the shape of the posterior distributions of the DM parameters (see \autoref{fig:Lshape}) confirms this hypothesis: with a poor frequency coverage (\texttt{DR2full}) there is a strong degeneracy between the DM parameters and the common red noise parameters, which is shown by the L-shape of the posterior distributions. When the coverage is improved (\texttt{DR2FC}), the long tails of the L-shape distributions disappear and the DM parameters are well constrained. In order to quantify the degeneracy of the CRN posterior distributions and the individual pulsars noise posterior distributions, we chose the following approach:
\begin{enumerate}
    \item for each individual noise, we select the 2D distribution of $\rm log_{10}A_{\rm DM/RN}$ and $\rm log_{10}A_{\rm CRN}$
    \item for each of these 2D posterior distributions, we generate a bivariate Gaussian distribution with same mean and covariance matrix; then we generate from that Gaussian distribution a number of samples equal to the 2D posterior distribution of interest
    \item we compare each generated Gaussian distribution with its parent posterior distribution by evaluating the Jensen-Shannon divergence \citep[JSD,][]{js} between the two samples. The JSD measures the similarity between two distributions and it is symmetric and bound between 0 and ln(2). If the JSD is high, it means that our 2D distribution is not well approximated by a bivariate Gaussian: this non-similarity is due to significant tails that characterise a degenerate, L-shaped distribution.
\end{enumerate}

The effectiveness of this test depends on the assumption that the posterior distributions are either covariate Gaussians or L-shaped distributions and never have bimodalities or other features that would make them differ from a Gaussian, but without any degeneracy between the parameters. The validity of this assumption is not trivial a priori but, for the cases analysed in this investigation, has been empirically verified by visual inspection of the posterior distributions. An example of the application of this method is shown in \autoref{fig:Lshape}. The \texttt{DR2full} degenerate distribution produces a JSD of 0.65 (let us recall that the maximum is $\sim$ 0.69 = ln(2)), while the non-degenerate posterior distribution obtained with \texttt{DR2FC} gives JSD = 0.19.

For each dataset realization, we have 31 individual pulsar noises, therefore we compute with this method 31 values of JSD. In order to extract only one number which is representative of how much degenerate the individual noises are with the common red noise, we chose to perform a weighted average of the JSD values. The weights are proportional to the contribution of each pulsar to the total GWB evidence. This contribution is evaluated exploiting the signal-to-noise ratio (SNR) defined in \cite{Rosado_2015}: 
\begin{equation}
    {\rm SNR}_B^2 = 2\sum_{a=1}^M\sum_{b>a}^M\frac{\Gamma_{ab}^2S^2(f)T_{ab}}{[P_a(f)+S(f)][P_b(f)+S(f)]+\Gamma_{ab}^2S^2(f)},
\end{equation}
where $P_a(f)$ and $P_b(f)$ are the noise spectral densities in the two pulsars $a$ and $b$, $\Gamma_{ab}$ is the angular correlation expected for a GWB produced by a population of SMBHBs, which is given by the Hellings and Downs curve in \autoref{eq:hd}. $S(f)$ is the power spectral density of the GWB signal, evaluated as 
\begin{equation}
S(f) = \biggl(\frac{h_c^2(f)}{12\pi^2f^3T_{\rm obs}}\biggr)^{1/2},
\end{equation}
with $h_c^2(f)$ given by \autoref{eq:spectrum}. Finally, $T_{ab}$ is the maximum time interval over which observations from pulsar $a$ and pulsar $b$ overlap. This formula evaluates the SNR produced by the entire array of pulsars, thus the contribution of a single pulsar $a$ is estimated as 
\begin{equation}
    {\rm SNR}_{B, a} = \sqrt{\sum_{b\neq a}^M\frac{\Gamma_{ab}^2S^2(f)T_{ab}}{[P_a(f)+S(f)][P_b(f)+S(f)]+\Gamma_{ab}^2S^2(f)}}.
\end{equation}
Therefore, the weighted average of the JSD values is obtained as 
\begin{equation}
    \overline{\rm JSD} = \frac{\sum_{i=1}^{i=31}{\rm JSD}_i\cdot {\rm SNR}_{B, i}}{\sum_{i=1}^{i=31}{\rm SNR}_{B, i}},
\end{equation}
where the index $i$ spans the individual noise processes and ${\rm SNR}_i$ is the contribution to the SNR of the pulsar responsible for the noise process.

\begin{figure}
   \centering
   \includegraphics[width=0.5\textwidth]{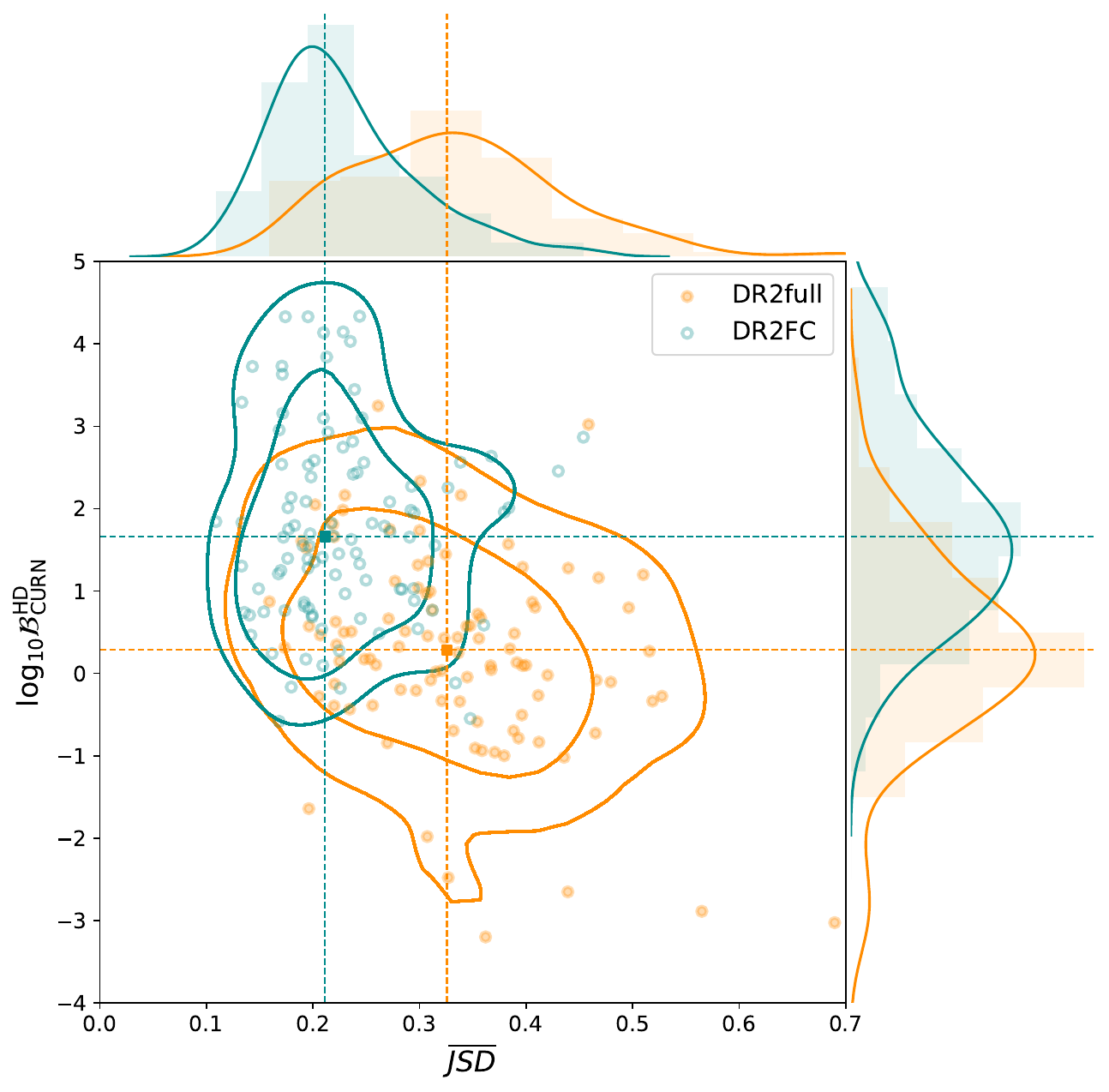}
   \caption{Scatter plot of the weighted average JSD (representing how degenerate the individual noise parameters are with the common red noise parameters) versus log$_{10}\mathcal{B}^{\rm HD}_{\rm CURN}$. The orange points are the 100 realizations of \texttt{DR2full}-like datasets, the green points are the 100 realizations of \texttt{DR2FC}-like datasets. Contours represent 2D 68$\%$ and 90$\%$ C.I. Histograms and smoothed distributions on the top and left side of the plot represent marginalised distributions of $\overline{{\rm JSD}}$ and log$_{10}\mathcal{B}^{\rm HD}_{\rm CURN}$ respectively.}
   \label{fig:bf_jsd}%
\end{figure}

\autoref{fig:bf_jsd} shows the scatter plot of the weighted averaged JSD versus the HD vs CURN Bayes factor for the 100 realizations of the two datasets. 
The distribution of ($\overline{\rm JSD}, \rm log_{10}\mathcal{B}^{\rm HD}_{\rm CURN}$) for the two datasets shows that \texttt{DR2full} results in lower values of log$_{10}\mathcal{B}^{\rm HD}_{\rm CURN}$ and higher values of $\overline{\rm JSD}$ with respect to \texttt{DR2FC}. We can therefore conclude that in the majority of the noise realisations the improved frequency coverage does indeed reduce the degeneracy of the noise parameters with the GWB parameters; and that this reduction in degeneracy is accompanied by an increase in the $\mathcal{B}^{\rm HD}_{\rm CURN}$, implying that there is a significant correlation between the $\mathcal{B}^{\rm HD}_{\rm CURN}$ and the degeneracy between the individual pulsar noise and the common red noise.

\subsubsection{Impact of the frequency coverage on the GWB parameter estimation}
\begin{figure}
   \centering
   \includegraphics[width=0.5\textwidth]{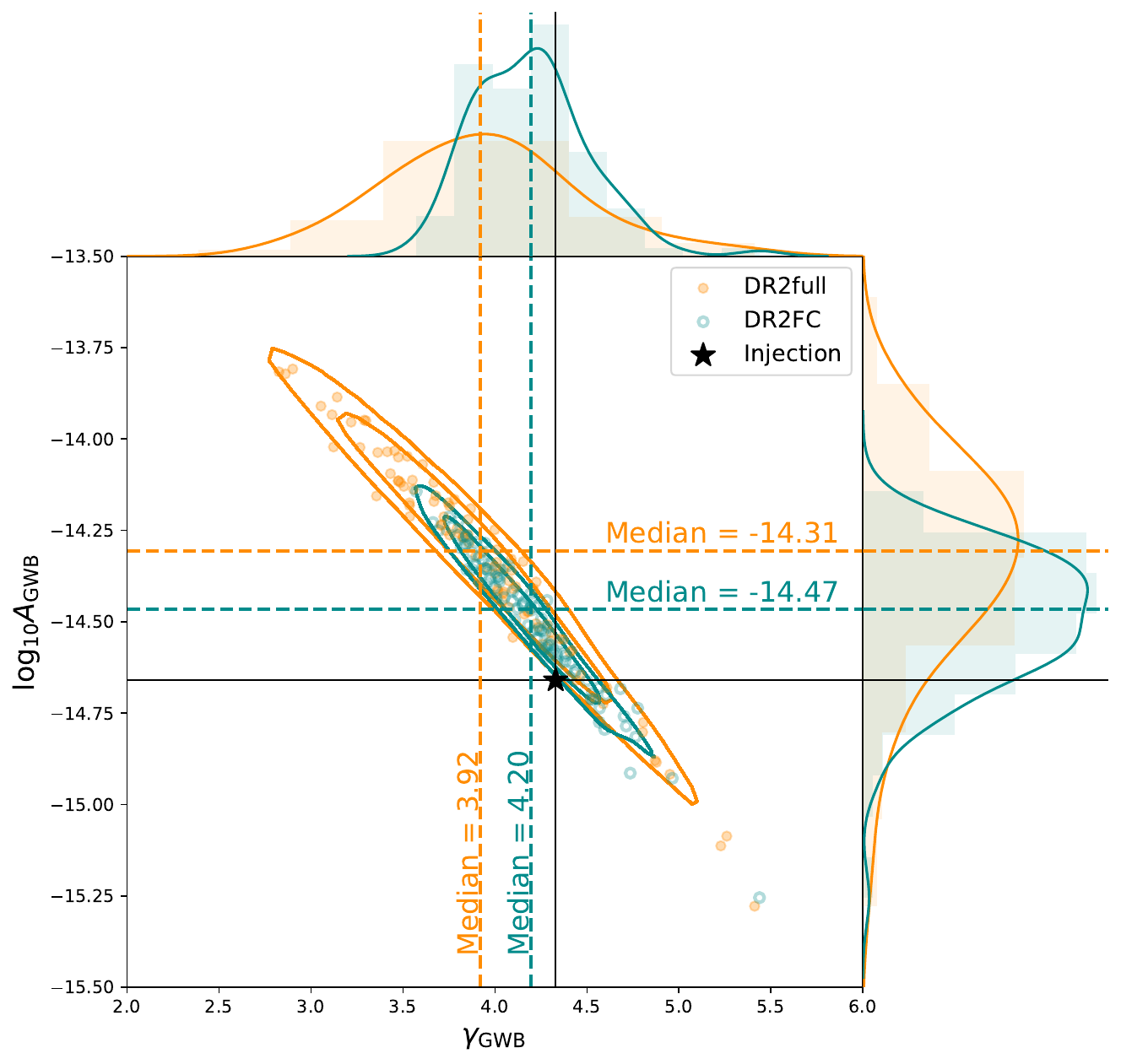}
   \caption{Median of the posterior distributions of log$_{10}A_{\rm GWB}$ and $\gamma_{\rm GWB}$ obtained with the 100 realisations of \texttt{DR2full} (orange) and \texttt{DR2FC} (green). Dashed lines show the median of the 1D distributions, and the black solid line is the injection. Histograms and smoothed distributions on the top and left side of the plot represent the marginalised posterior distributions of the parameters recovery.}
   \label{fig:rec}%
\end{figure}
We now use the same two datasets -- \texttt{DR2full} and \texttt{DR2FC} -- to investigate the effect of frequency coverage on the recovery of the GWB parameters.
The results obtained with the 100 realisations of \texttt{DR2full} are, quoting the median and the 68$\%$ C.I., log$_{10}A_{\rm GWB}=-14.31^{+0.26}_{-0.26}$ for the amplitude and $\gamma_{\rm GWB}=3.92_{-0.47}^{+0.46}$ for the slope of the GWB. Therefore, as for the Bayes factor, the amplitude and slope obtained from the real \texttt{DR2full} analysis \citep[log$_{10}A_{\rm GWB}=-14.54^{+0.28}_{-0.41}$ and $\gamma_{\rm GWB}=4.19_{-0.63}^{+0.73}$,][]{wm3} are within the 68$\%$ C.I. obtained from our simulated datasets. From \autoref{fig:rec} it can be seen that the recovered GWB is on average higher and flatter than the injected background -- log$_{10}A_{\rm GWB}\sim-14.66$, $\gamma_{\rm GWB}\sim4. 33$ -- and the injected values are not even on the eigen-diagonal of the distribution, but slightly below it; in fact, the injected value is not within the 90$\%$ C.I. of the 2-D posterior distribution, and the recovered amplitude is on average higher than the injected value.

\begin{figure}
   \centering
   \includegraphics[width=0.5\textwidth]{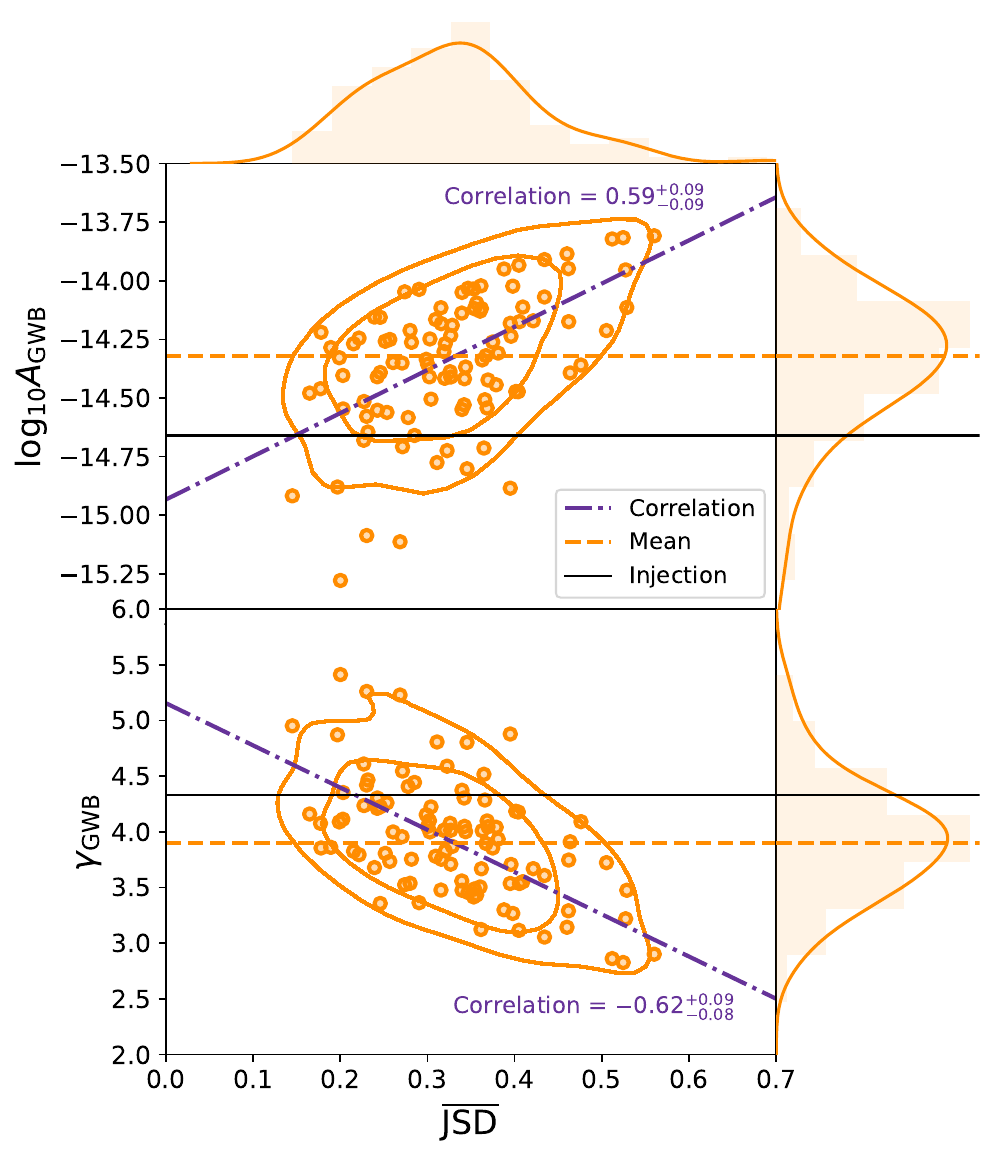}
   \caption{Weighted average JSD versus GWB amplitude ({\it top panel}) and the GWB slope ({\it bottom panel}) for the 100 simulated \texttt{DR2full} datasets. 
    In each panel, dashed lines correspond to the mean of the distributions, dotted dashed lines are the eigen-diagonals of the distributions, showing strong correlation between the weighted average JSD and the recovered value of the two GWB parameters. Black solid horizontal lines are the injected values (taking in mind that the signal injected has not a perfect power-law spectrum). Histograms and smoothed distributions on the top and left side of the plot represent the marginalised posterior distributions of the corresponding quantities.}
   \label{fig:dr2full_rec}%
\end{figure}

\autoref{fig:dr2full_rec} shows the weighted average JSD and the recovery of the GWB parameters for the 100 realisations of \texttt{DR2full}. For both parameters, there is a correlation between the recovered value and the average JSD with a significance greater than 5$\sigma$. It is a positive correlation in the case of the GWB amplitude, $\rho = 0.59^{+0.09}_{-0.09}$, and a negative correlation in the case of the GWB slope, $\rho = -0.62^{+0.08}_{-0.09}$. This means that the more degenerate the individual noise parameters are with the GWB parameters, the flatter and higher is the recovered background with respect to the injection. Note, however, that even at low values of $\overline{\rm JSD}$, where the L-shape is not significant, the distribution of the recovered parameters is not centred on the injected values. This means that the degeneracy is not the only factor responsible for the bias in the recovery of the GWB parameters. This is consistent with the fact that the bias in the recovery is not solved by the improvement in the frequency coverage: the recovered values obtained with \texttt{DR2FC}, indeed, are also clustered at higher amplitude values and lower slope values with respect to the injection (see \autoref{fig:rec}). Nevertheless, the bias is slightly reduced in this case, with respect to \texttt{DR2full}. The main difference between the recoveries obtained with the two datasets, which can be seen in \autoref{fig:rec}, is the width of the distributions: the one obtained with \texttt{DR2FC} is much narrower because the noise is better modelled in this case and therefore there is less dependence on the particular realisation of noise.

The residual bias is probably caused by the mismatch between the power-law template used in the recovery and the spectrum of the real injected population, as investigated in \cite{2024A&A...683A.201V}. Indeed, the minimum of the EPTA DR2 sensitivity corresponds to the first two bins, which are therefore expected to contribute most to the recovery. As can be seen in \autoref{fig:GWB_spectrum}, even though the whole spectrum is well described by a $f^{-2/3}$ power-law, the first two bins seem to slightly deviate from this slope, corresponding to a higher and flatter spectrum.

\subsection{Understanding the RN-DM degenerate likelihood : a toy model}

In order to gain a better inside on the results obtained in the previous section, we build a toy model to mathematically describe how the RN-DM degeneracy affects the PTA likelihood function and the recovery of the GWB signal.
Consider a pulsar with timing residuals $\delta \vec{t}$ containing only white noise, RN and DM noise. The pulsar was observed for $N_E$ epochs with $N_\nu$ frequency channels per epoch. For simplicity, we treat the RN and DM as single frequency deterministic signals to allow straightforward maximum likelihood calculation. We define the RN basis $F$ and the DM basis $F_{\rm DM}$ at times $\vec{t}$ and frequencies $\vec{\nu}$ as

\begin{equation}
    F = [\sin(2\pi f \vec{t}), \cos(2\pi f \vec{t})]\,\,\,\,\,\,\,;\\
    F_{\rm DM} = \left (\frac{\nu_{\rm ref}}{\vec{\nu}}\right)^2 F,
\end{equation}
so that
\begin{equation}
\begin{aligned}
    &F^\top F  = \frac{N_E}{2} N_\nu \delta_{ij},\\
    &F_{\rm DM}^\top F_{\rm DM}  = \frac{N_E}{2} \sum_a \left ( \frac{\nu_{\rm ref}}{\nu_a}\right )^4 \delta_{ij} = \frac{N_E}{2} N_\nu \bar{\nu}_4 \delta_{ij},\\
    &F^\top F_{\rm DM}  = \frac{N_E}{2} \sum_a \left ( \frac{\nu_{\rm ref}}{\nu_a}\right )^2 \delta_{ij} = \frac{N_E}{2} N_\nu \bar{\nu}_2\delta_{ij}.
\end{aligned}
\end{equation}

For a pulsar with RN, DM and white noise, the timing residuals are 

\begin{equation}
    \delta t = F \cdot c_{\rm RN} + F_{\rm DM} \cdot c_{\rm DM} + n,
\end{equation}
where $n \sim \mathcal{N}(0, \sigma)$ is the Gaussian white noise, and $c_{\rm RN}$ and $c_{\rm DM}$ are the true values for RN and DM amplitude respectively. In the following, we consider that we are in the red noise dominated regime so we can neglect white noise $n$ in the timing residuals. We write the likelihood :

\begin{equation}
    \ln \mathcal{L} \propto -\frac{1}{2\sigma^2} ||\delta t - (F \cdot a_{\rm RN} + F_{\rm DM} \cdot a_{\rm DM})||^2
\end{equation}
where $||y||^2 = y^\top y$. We can then investigate two separate situations.

\subsubsection{The completely degenerate case}

\begin{figure}
   \centering
   \includegraphics[width=0.5\textwidth]{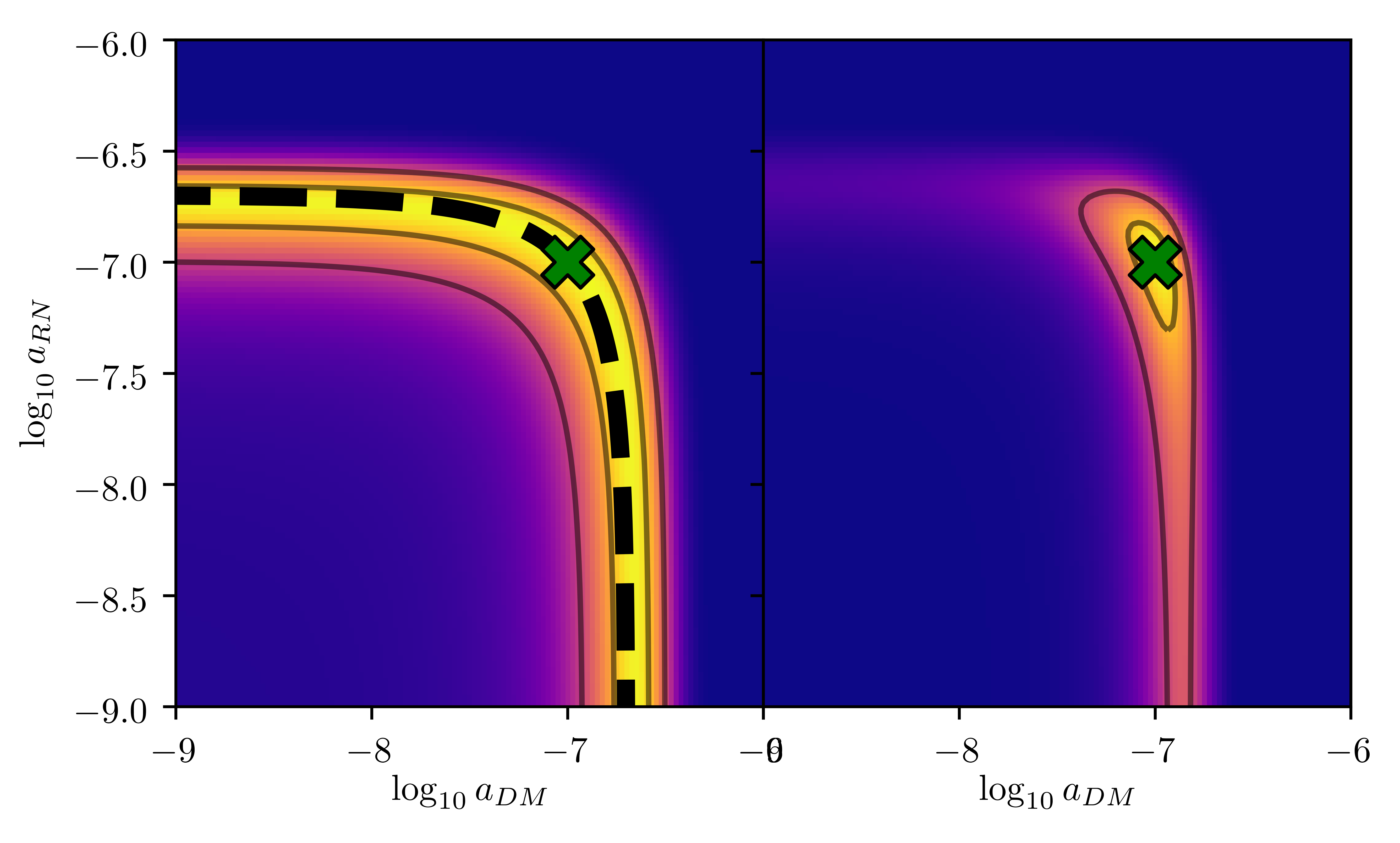}
   \caption{Example of degenerate likelihood. The true value of the parameters is marked with the green cross. Left panel: completely degenerate case $\nu = \nu_{\rm ref}$, in which RN and DM cannot be distinguished. The dashed line shows the degenerate maximum likelihood behaving as $y \propto 1-x$ in log-scale. Right panel: same likelihood with $\nu \neq \nu_{\rm ref}$. The degenerate tails are less pronounced and the maximum likelihood lies around the true value of parameters.}
   \label{fig:l_shaped_likelihood}
\end{figure}

If all $\nu = \nu_{\rm ref}$, we have $F = F_{\rm DM}$, which yields

\begin{equation}
     \ln \mathcal{L} \propto -\frac{1}{2\sigma^2} ||F \cdot (c_{\rm RN} + c_{\rm DM} - a_{\rm RN} - a_{\rm DM})||^2.
\end{equation}
In that case, $a_{\rm RN}$ and $a_{\rm DM}$ are completely degenerate parameters of the model since they have the same effect on the likelihood. Consequently, trying to find their maximum likelihood value gives an infinite number of solutions and they cannot be constrained. The maximum likelihood lies around the line where $a_{\rm RN} = x(c_{\rm RN} + c_{\rm DM})$ and $a_{\rm DM} = (1-x)(c_{\rm RN} + c_{\rm DM})$, where $x=[0, 1]$ is a real number. This produces the typical L-shaped distribution that we find on \autoref{fig:l_shaped_likelihood}.

\begin{figure}
   \centering
   \includegraphics[width=0.5\textwidth]{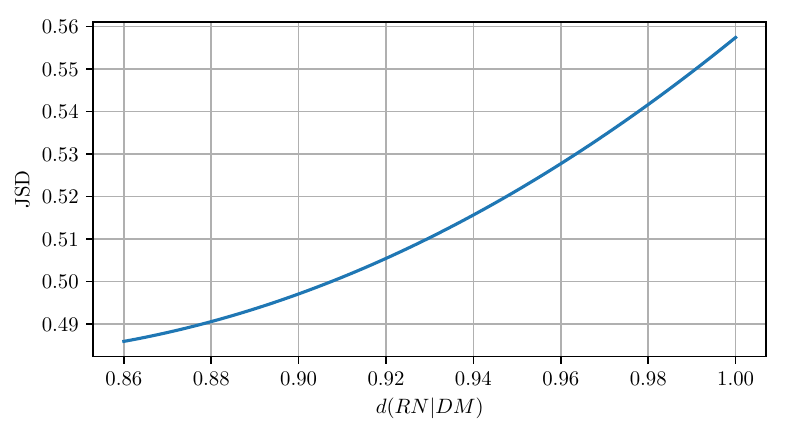}
   \caption{JSD as a function of $d({\rm RN|DM})$ estimated using the toy model.}
   \label{fig:JSD_vs_d}
\end{figure}

The left tail corresponds to the regime where $a_{\rm RN}$ accounts for all the noise ($x \approx 1)$ while $a_{\rm DM} \approx 0$ and the right tail corresponds to the regime where $a_{\rm DM}$ accounts for all the noise ($x \approx 0)$ while $a_{\rm RN} \approx 0$.

\subsubsection{The non-degenerate case}

If $\nu \neq \nu_{\rm ref}$, we have $F \neq F_{\rm DM}$, which yields

\begin{equation}
     \ln \mathcal{L} \propto -\frac{1}{2\sigma^2} ||F \cdot (c_{\rm RN} - a_{\rm RN}) + F_{\rm DM} \cdot (c_{\rm DM} - a_{\rm DM})||^2.
\end{equation}
Now that $F$ and $F_{\rm DM}$ are disentangled, the maximum likelihood values of parameters $a_{\rm RN}$ and $a_{\rm DM}$ are the true values $c_{\rm RN}$ and $c_{\rm DM}$. Nevertheless, the tails of the previously introduced L-shape do not disappear. In fact, if the distribution of frequencies $\nu_a$ is very narrow and centered on the reference frequency $\nu_{\rm ref}$, we are still in the degenerate case. We estimate the significance of each tail by setting $a_{\rm RN}=0$ or $a_{\rm DM}=0$ and finding the maximum likelihood for the remaining free parameters $\hat{a}_{\rm RN}, \hat{a}_{\rm DM}$. We have
\begin{equation}
\begin{aligned}
\hat{a}_{\rm RN} & = c_{\rm RN} + \bar{\nu}_2 c_{\rm DM} \\
\hat{a}_{\rm DM} & = \frac{\bar{\nu}_2}{\bar{\nu}_4} c_{\rm RN} + c_{\rm DM}
\end{aligned}
\end{equation}
and the corresponding maximum likelihoods
\begin{equation}
\begin{aligned}
\ln \mathcal{L}(\delta t | \hat{a}_{\rm RN}) & \propto -\frac{N_\nu}{2\sigma^2} \left [ \bar{\nu}_4 - \bar{\nu}^2_2 \right ] c_{\rm DM}^2 = -\frac{N_\nu}{2\sigma^2} \sigma^2_\nu c_{\rm DM}^2 \\
\ln \mathcal{L}(\delta t | \hat{a}_{\rm DM}) & \propto -\frac{N_\nu}{2\sigma^2} \left [ 1 - \frac{\bar{\nu}_2^2}{\bar{\nu}_4} \right ] c_{\rm RN}^2 = -\frac{N_\nu}{2\sigma^2} \sigma^2_\nu \frac{c_{\rm RN}^2}{\bar{\nu}_4}.
\end{aligned}
\end{equation}

The significance of each tail depends on the amplitude of the noise component it is trying to fit and scales as the variance $\sigma_\nu$ of the inverse squared frequencies $\nu$ per epoch. We can isolate from the previous expressions a coefficient that is only function of the frequency coverage and quantifies the degeneracy between RN and DM
\begin{equation}
    d({\rm RN|DM}) = \exp \left \{-\frac{1}{2} N_\nu \sigma_\nu ^2 \right \}
\label{eq:degen}
\end{equation}
that is bound between 0 and 1. If $\nu = \nu_{\rm ref}$, $d({\rm RN|DM})=1$, if not, $0 < d({\rm RN|DM}) < 1$.

In \autoref{fig:JSD_vs_d}, we verify the relationship between the coefficient $d({\rm RN|DM})$ and the JSD metric defined in section \ref{BF_JSD} characterising the degeneracy from the posterior distribution. For decreasing values of $d({\rm RN|DM})$, we have decreasing values of JSD.

\subsubsection{Parameter estimation and SNR}

We showed that the degeneracy between RN and DM will bias parameter estimation. In the weak signal regime, the optimal SNR for a GWB with HD correlations is \citep{2018PhRvD..98d4003V}

\begin{equation}
    \rho_{HD} = \frac{\sum_{ab} \delta t_a^\top C_a ^{-1} S_{ab} C_b^{-1} \delta t_b}{\left [ \sum _{ab} \textrm{tr} \left ( C_a ^{-1} S_{ab} C_b^{-1} S_{ba} \right ) \right]^{1/2}},
    \label{eq:os_snr}
\end{equation}
where $\Gamma_{ab}$ is the Hellings-Downs correlation coefficient between pulsars $a$ and $b$, $S_h$ is the PSD of the GWB signal and $S_a, S_b$ are the PSD of the noise in pulsars $a$ and $b$. It is clear that an erroneous estimation of signal and noise parameters will affect the SNR.

To evaluate the influence of frequency coverage on SNR estimation, we compute the likelihood for a toy dataset of 25 pulsars containing white noise, DM noise and a GWB in their residuals with 10 years of observation and a 2 weeks cadence. The GWB and DM noise spectra are modelled as a power-law with fixed spectral index $\gamma=3$ and same amplitude. Since the level of noise is identical in each pulsar ($A_{\rm CRN}=A_{{\rm DM},a}=10^{-14}$ and $\sigma = 10^{-6}s$), we assume a factorised likelihood form \citep{Taylor_2022} where we only vary the amplitudes of CRN and DM as :

\begin{equation}
    \ln \mathcal{L}(\delta t|\Vec{\theta}) = \sum _{a=1} ^{N_{\rm psr}} \ln \mathcal{L}(\delta t|\Vec{\theta}_{\rm CRN}, \Vec{\theta}_{{\rm DM},a}) 
\end{equation}
where $\Vec{\theta}_{\rm CRN}$ are the common red noise parameters and the $\Vec{\theta}_{{\rm DM},a}$ are the DM noise parameters for pulsar $a$.

In this simplified configuration, we consider each $\ln \mathcal{L}(\delta t|\Vec{\theta}_{\rm CRN}, \Vec{\theta}_{{\rm DM},a})$ to be identical and estimate it once on a 2D grid using the expression given in \autoref{eq:likelihood}. Thus, for a given prior $\pi(\vec{\theta})$, we can directly produce sets of samples by first drawing $\Vec{\theta}_{\rm CRN}$ from the 1D marginalised posterior distribution $p(\Vec{\theta}_{\rm CRN}|\delta t)$ and independently drawing the $\Vec{\theta}_{{\rm DM},a}$ from the conditional distribution $p(\Vec{\theta}_{{\rm DM},a}|\delta t, \Vec{\theta}_{\rm CRN})$. We generate many fake PTAs for which we only varied the frequency coverage, splitting the TOAs into 5 different frequency channels distributed in the L-band, i.e. $\nu = 1-2$ GHz. For each simulation, we compute the marginalised OS SNR using \autoref{eq:os_snr} and the degeneracy coefficient $d({\rm RN|DM})$ using \autoref{eq:degen}. 

\begin{figure}
   \centering
   \includegraphics[width=0.5\textwidth]{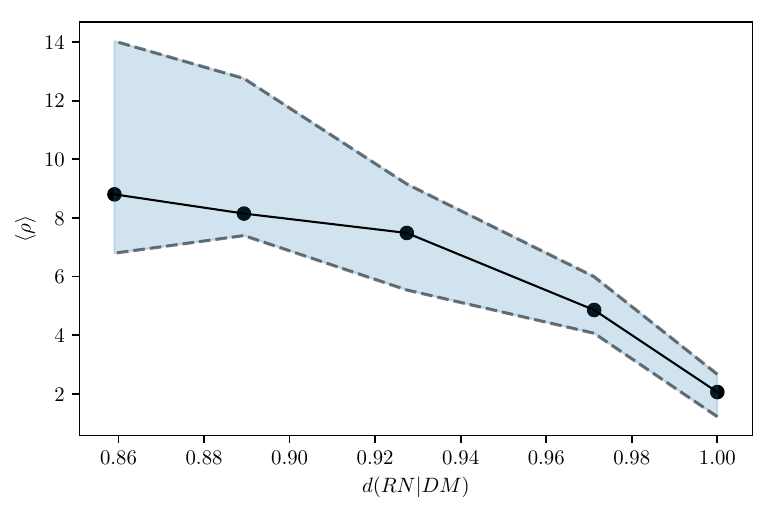}
   \caption{SNR as a function of the degeneracy coefficient from 1000 simulations. The blue region represents the 1-$\sigma$ credible region.}
   \label{fig:snr_degen}
\end{figure}

\begin{figure}
   \centering
   \includegraphics[width=0.5\textwidth]{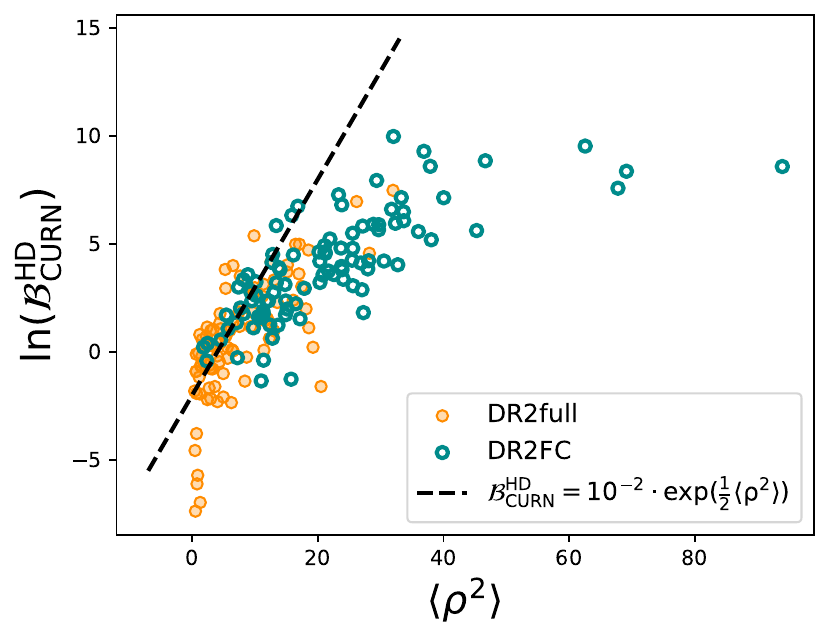}
   \caption{SNR and ln($\mathcal{B}$) for the 100 realizations of \texttt{DR2full} (orange) and \texttt{DR2FC} (green). The dashed line represent the expected relation in the weak signal regime.} 
   \label{fig:bf_snr}
\end{figure}

The recovered SNRs as a function of $d({\rm RN|DM})$ is shown in \autoref{fig:snr_degen}. The SNR shows a clear dependence on the frequency coverage. Adding higher frequency channels in our array reduces the amplitude of DM and improves the SNR (since DM varies as as $\nu^{-2}$), while adding lower frequency channels gives a better characterisation of the latter because it is more pronounced at low $\nu$. We can see that even for slightly improved coverage, the difference in SNR can be significant. Since in the weak signal regime, we can link the SNR to the Bayes factor through the formula \citep{Vallisneri_2023}
\begin{equation}
    \ln \mathcal{B}^{\rm HD}_{\rm CURN} \simeq \frac{1}{2} \langle \rho \rangle^2
\end{equation}
a slight increase in the coverage, and therefore in the SNR, leads to an exponential growth of the Bayes factor (see \autoref{fig:bf_snr}).

The results from the simple simulations described in this section can be compared with those from the realistic simulations: we can evaluate the HD SNR given by the 100 realizations of the two realistic datasets \texttt{DR2full} and \texttt{DR2FC} with the noise marginalised optimal statistic method \citep{2018PhRvD..98d4003V}. We can also evaluate the degeneracy coefficient $d({\rm RN|DM})$, which depends only on the frequency coverage $\sigma_{\nu}$, for the two realistic datasets. The values of HD SNR averaged over the noise realizations are respectively $\sim 2.50$ and $\sim 4.42$, while the values of $d({\rm RN|DM})$ are $\sim 0.97$ and $\sim 0.95$. The average SNR $\langle\rho\rangle$ shown in \autoref{fig:snr_degen} was obtained for an ideal PTA with evenly distributed data, resulting in higher values compared to realistic simulations. Nevertheless, we can compare the relative increase in the average SNR that follows an improvement in the frequency coverage from $d({\rm RN|DM}) = 0.97$ to $d({\rm RN|DM}) = 0.95$ (i.e. from \texttt{DR2full} to \texttt{DR2FC}). The ratio between \texttt{DR2FC} HD SNR and \texttt{DR2full} HD SNR is $\sim 4.43/2.50\sim1.77\pm0.91$, while the ratio extrapolated from the $\langle\rho\rangle$ vs $d({\rm RN|DM})$ curve is $6.16/4.80\sim1.29\pm0.78$. Therefore, they are very well compatible. Still, the good agreement is accompanied by large errors in the two estimates due to the dependence on the noise realisation.

Even slight changes in the frequency coverage of a PTA can have a strong impact on the significance of the recovered signal. It is then fundamental to have an accurate characterisation of the DM noise to remove potential degeneracies between DM and RN. Next generation of observatories at very low frequency where DM noise is high, such as SKA \citep[e.g.][]{Bhat_2018}, LOFAR \citep[e.g.][]{Kondratiev_2016}, Nenufar \citep{Bondonneau_2021} and Chime \citep{Amiri_2021}, will provide good estimates of DM to better distinguish the latter from achromatic noise \citep{signore_dottore_iraci}.

\section{Conclusions}
\label{sec:con}

Noise modelling is a cornerstone of a functioning PTA, and the importance for customised noise models for each pulsar was highlighted in previous studies \citep{Chalumeau_2021}. In this article, we have explored the degeneracies that may occur between the noise models and the GW signal models. In particular, we focused on the DM noise that has a chromatic dependence on the incoming photon's frequency. The characterisation of the latter requires multi-frequency observation capabilities at the radio-telescope, since having a single frequency channel of observation makes it impossible to differentiate between achromatic noise and chromatic noise.

We simulated realistic PTA datasets based on the second data release of the EPTA collaboration. This dataset is dominated by a single frequency of observation in its first half, possibly introducing biases in the estimates of the GWB signal and DM noise. We produced two versions of this dataset, (i) with the real frequency channel distribution and (ii) with a homogeneous frequency channel distribution. For each of the two datasets we performed 100 simulations with the same realization of GWB signal, but 100 different realizations of noise. We found that in 91\% of the cases, the Bayes factor in favour of a GWB was significantly boosted by the improvement in the frequency coverage, as shown in \autoref{fig:BF_fc}, with an average value 20 times higher compared to the datasets mimicking EPTA DR2. This is because without proper frequency coverage, DM can easily be mis-modeled, leaking into the RN budget. Since DM affects all pulsars and can be orders of magnitude higher than RN, by leaking in the red noise budget it can give rise to a spurious common red signal that can severely interfere with the GWB recovery.
The degeneracy that exists between DM and RN parameters can be observed in the 2D posterior probability distributions (see \autoref{fig:Lshape} and \autoref{fig:l_shaped_likelihood}). We proposed a metric to measure the amount of degeneracy based on the Jensen Shannon divergence and show that a reduced level of degeneracy is correlated with an improved significance of the GWB (\autoref{fig:bf_jsd}). Indeed, since parameter estimation is strongly affected by the observation frequency coverage, the SNR and the Bayes factor in favour of a GWB will deteriorate as the coverage decreases (see \autoref{fig:snr_degen}). As expected, the recovery of the GWB parameters is also strongly correlated with the degeneracy between the DM and GWB parameters. In particular, a larger degeneracy is associated with a bias of the GWB recovery towards a flatter and higher spectrum.

In conclusion, this work shows that it is not particularly surprising that a dataset with the characteristics of EPTA DR2full does not return a significant detection of a GWB with a nominal amplitude of 2.4$\times 10^{-15}$ \cite{wm3}, with one of the reasons being the highly inhomogeneous frequency coverage. It also shows that care must be taken when forecasting future PTA capabilities or when devising observing strategies to maximize the scientific potential of a PTA. Not properly taking into account the DM-RN degeneracy can lead to overoptimistic results and biased forecasts. Finally, it is of paramount importance to include in the PTA datasets data from the new generations of radio telescopes (LOFAR, Chime, SKA) with increased frequency band, to help distinguishing chromatic features from GW signals. Understanding the DM noise and all other chromatic features that are present in pulsar timing data is a fundamental aspect of PTAs that should be one of the priorities for the future.

\begin{acknowledgements}
M.F. acknowledges financial support from MUR through the PRIN 2022 project "General Relativistic Astrometry and Pulsar Experiment" (GRAPE, 2022-7MYL2X).
A.S. acknowledges financial support provided under the European Union’s H2020 ERC Consolidator Grant ``Binary Massive Black Hole Astrophysics'' (B Massive, Grant Agreement: 818691).
\end{acknowledgements}

\bibliographystyle{aa}
\bibliography{bibliography}

\end{document}